# Imaging Microwave and DC Magnetic Fields in a Vapor-Cell Rb Atomic Clock

C. Affolderbach, G.-X. Du, T. Bandi, A. Horsley, P. Treutlein, and G. Mileti

## ABSTRACT

We report on the experimental measurement of the DC and microwave magnetic field distributions inside a recently-developed compact magnetron-type microwave cavity, mounted inside the physics package of a high-performance vapor-cell atomic frequency standard. Images of the microwave field distribution with sub-100 µm lateral spatial resolution are obtained by pulsed optical-microwave Rabi measurements, using the Rb atoms inside the cell as field probes and detecting with a CCD camera. Asymmetries observed in the microwave field images can be attributed to the precise practical realization of the cavity and the Rb vapor cell. Similar spatially-resolved images of the DC magnetic field distribution are obtained by Ramsey-type measurements. The $T_2$ relaxation time in the Rb vapor cell is found to be position dependent, and correlates with the gradient of the DC magnetic field. The presented method is highly useful for experimental in-situ characterization of DC magnetic fields and resonant microwave structures, for atomic clocks or other atom-based sensors and instrumentation.

INDEX TERMS:

Atomic clocks, Diode Lasers, Microwave measurements, Microwave resonators, Microwave spectroscopy, Optical Pumping.

This work was supported in part by the Swiss National Science Foundation (SNFS grant no. 149901, 140712 and 140681) and the European Metrology Research Programme (EMRP project IND55-Mclocks). The EMRP is jointly funded by the EMRP participating countries within EURAMET and the European Union.

T. Bandi, C. Affolderbach and G. Mileti are with the Laboratoire Temps-Fréquence (LTF), Institut de Physique, Université de Neuchâtel, Neuchâtel, Switzerland (e-mail: gaetano.mileti@unine.ch). T. Bandi is now at Quantum Sciences and Technology Group (QSTG), Jet Propulsion Laboratory (JPL), California Institute of Technology, Pasadena, USA.

G.-X. Du, A. Horsley, and P. Treutlein are with Departement Physik, Universität Basel, Switzerland. (e-mail: philipp.treutlein@unibas.ch).

## I. INTRODUCTION

Compact vapor-cell atomic frequency standards (atomic clocks) [1], [2] are today widely used in applications such as telecommunication networks [3] or satellite navigation systems [4], and are also of interest for other scientific or industrial applications. In view of future demand for highly compact but nevertheless high-performance vapor-cell clocks in these fields, laboratory clocks with state-of-the-art fractional clock frequency stabilities of $1.4\times10^{-13}$ at an integration time of $\tau$ = 1 second [5] and down to few $10^{-15}$ for $\tau = 10^4$ s [6] have been reported recently. Precise knowledge of the microwave and DC magnetic field distributions in such clocks is a key requirement for their development.

In a Rb vapor-cell atomic clock, the frequency of a quartz oscillator is stabilized to the frequency of the so-called microwave hyperfine "clock transition" $|F=1, m_F=0\rangle \leftrightarrow |F=2, m_F=0\rangle$ in the $5S_{1/2}$ ground state of $^{87}$Rb (at $\nu_{Rb}$ = 6'834'682'610.904'312 Hz [7], see Fig. 2c), generally detected using the optical-microwave double-resonance (DR) scheme. In Rb clocks based on the continuous-wave (cw) DR interrogation scheme [2], [5] a ground-state polarization is created by optical pumping [8] with a Rb lamp or laser, and the clock transition frequency is detected via a change in light intensity transmitted through the cell. In this scheme, the optical and microwave fields are applied simultaneously and continuously, which can cause perturbations of the clock frequency due to the light shift effect [5], [9]. In the pulsed interaction scheme [6], first a resonant laser pump pulse creates the ground state polarization, followed by two time-separated microwave pulses in the Ramsey scheme [10]. The atomic response is then read out by a laser detection pulse. With the optical and microwave interaction separated in time, the light shift effect can be significantly reduced in this scheme. In both approaches, the Rb atomic sample is held in a sealed vapor cell usually also containing a buffer gas to avoid Rb collisions with the cell walls, and the microwave is applied to the atoms using a microwave cavity resonator, which allows realizing compact clock physics packages. For all atomic clocks employing microwave cavity resonators to apply the microwave radiation to the atoms, the uniformity and homogeneity of the resonant microwave magnetic field inside the cavity is of critical importance for achieving strong clock signals: Selection rules for the clock transition require a microwave magnetic field oriented parallel to the quantization axis defined by the applied DC magnetic field, and the pulsed scheme in addition requires a homogeneous microwave field amplitude for applying $\pi/2$-pulses to all sampled atoms.

The design and study of different types of microwave resonator cavities with well-defined field distributions have been addressed, e.g., for compact vapor cell atomic clocks [11] and H-masers [12], high-performance Rb clocks based on the continuous-wave (cw) [13], [14] or pulsed optical pumping (POP) approach [6, 15], or miniaturized Rb clocks [16], [17]. In primary atomic fountain clocks, effects such as distributed phase shifts in the cavity can become relevant [18]. The impact of the microwave field distribution on Rb atomic clocks using buffer-gas and wall-coated cells has been discussed in [19], and effects like the microwave power shift have been shown to depend on the degree of inhomogeneity of the DC magnetic field (so-called "C-field") applied to the atomic sample [11], [20]. For a thorough understanding of a cell clock's performance limitations, knowledge of the precise distribution and homogeneity of both the DC and microwave magnetic fields applied to the atomic vapor is therefore of crucial importance.

In practice, it is difficult to experimentally measure the microwave magnetic field geometry and distribution in the cavity's final configuration, notably due to the presence of the vapor cell that prevents the placement of a field probe within its volume, and due to the field perturbations caused by the presence of such a probe. Most studies therefore rely on detailed analytic and/or numerical simulations of the field distribution, while the experimentally accessible parameters are generally integrated over the entire cell or cavity volume, e.g. by measuring S-parameters, resonance frequencies, quality factor, etc. [12], [13], [15]. Furthermore, due to the limited fabrication tolerances of glass-made vapor cells, the microwave field distribution inside the cavity will vary slightly from one cell to another, which can impact on the atom interrogation without necessarily being detectable by the methods mentioned above.

In this present work we exploit an imaging technique using the Rb atoms in our clock's vapor cell as local field probes [21], [22], [23] to obtain images of the microwave field distribution inside the cavity, on a fully assembled clock physics package and under real operating conditions, with a sub-100 µm lateral spatial resolution. We also show that a variant of this imaging technique can be used to obtain images of the DC magnetic field (C-field) applied to the cell. This allows assessing eventual changes in the static magnetic field even after extended clock operation time, without need to disassemble the clock physics package.

II. EXPERIMENTAL SETUP AND METHODS

In this section we describe the microwave cavity and clock physics package studied, as well as the experimental setup and detection schemes used.

*A. The Microwave Cavity*

The cavity under study here has been described in detail in previous publications [5], [13]. It is based on the loop-gap-resonator approach [24], also known as magnetron-type cavity [11]. In this cavity, a set of six electrodes placed inside a cylindrical electrically conducting cavity enclosure is used to create a resonance at the precise clock transition frequency (see Fig. 1a). This design also imposes a TE011-like mode geometry of the microwave field, with the magnetic field vector essentially parallel to the z-axis across the Rb cell, for effectively driving the clock transition (magnetic dipole transition). The cavity is highly compact, with an outer diameter and length of 40 mm and 35 mm, respectively, thus realizing a highly homogeneous microwave field across the cell using a cavity with a size below the clock transition's microwave wavelength. A photograph of the cavity with the Rb cell mounted inside is shown in Fig. 1b.

*B. Clock Physics Package and Imaging Setup*

For our field imaging studies we use the modified Rb atomic clock setup shown in Fig. 2a. The microwave cavity (as described in section II-A above) is placed inside a clock physics package also containing a thermostat, magnetic shields, and a solenoid (40 mm radius and 48 mm length) placed around the cavity for applying the DC magnetic field of $B_{dc} \approx 40$ µT, oriented parallel to the z axis. The vapor cell is made of borosilicate glass, has a diameter and length of 25 mm each, and is equipped with a cylindrical stem serving as Rb reservoir (see Fig. 1b). It contains isotopically enriched $^{87}$Rb, and 26 mbar buffer-gas mixture of $N_2$ and Ar for suppressed temperature sensitivity of the clock transition. The main cell body is held at a temperature of 55°C, and the stem at 40°C. Under these conditions we find a mean free path of $\lambda_{mf}$ = 5 µm for the Rb atoms in the cell.

Optical pumping and detection is achieved by a laser diode, whose frequency is stabilized to the $F_g = 2 \rightarrow F_e = 2, 3$ crossover transition of the Rb D2 line, observed in an auxiliary Rb cell without buffer gas, corresponding approximately to the center of the collisionally broadened and shifted optical transition in the buffer-gas cell [25]. The laser intensity incident to the cavity cell was 19.7 mW/cm$^2$, and was switched on and off using an acousto-optical modulator (AOM). The microwave radiation was produced by a laboratory microwave synthesizer, with a frequency close to the ≈ 6.835 GHz clock transition frequency, and its pulses controlled by switches. By setting the microwave frequency to values corresponding to the different resonances $i$ = 1 to 7 shown in Fig. 2b, the different Zeeman components of the Rb hyperfine ground-state transition indicated in Fig. 2c can be selected. The typical microwave power level injected into the cavity is around +22 dBm. The laser light level transmitted through the vapor cell is then mapped onto a CCD camera or a photodiode, using imaging optics.

C. Detection Schemes

Field imaging is performed using the method presented in [23], based on pulsed interaction schemes using Rabi measurements and Ramsey measurements. In both schemes a first "pump" laser pulse depopulates the $F_g$ = 2 ground-state level, followed by one or two microwave pulses for coherent interaction, and finally a weak probe laser pulse reads out the resulting atomic population in the $F_g$ = 2 state. The variation in optical density $\Delta OD$ of the atomic sample induced by the pulsed interaction scheme is then calculated from the laser intensities transmitted through the cell as described in [23]. While in [23] imaging was performed on a micro-fabricated vapor cell resulting in a spatial resolution in z-direction defined by the cell thickness of 2 mm, we here apply the imaging technique to a thick cell of 25 mm length.

Rabi measurements, using *one* microwave pulse of variable duration $dt_{mw}$ between the optical pump and probe pulses, were employed for imaging of the microwave magnetic field amplitudes in the cavity. In this case, the microwave frequency is tuned to the center of the selected Zeeman component of the hyperfine transition. The observed variation in optical density as function of $dt_{mw}$ shows Rabi oscillations and is described by

$$\Delta OD = A - B \exp\left(-\frac{dt_{mw}}{\tau_1}\right) + C \exp\left(-\frac{dt_{mw}}{\tau_2}\right) \sin(\Omega\, dt_{mw} + \phi) \qquad (1)$$

Here the fit parameters are an overall constant offset $A$ in optical density, amplitude $B$ of the relaxation of population with its related time constant $\tau_1$, and the amplitude $C$, time constant $\tau_2$, Rabi frequency $\Omega$, and phase $\phi$ of the Rabi oscillations introduced by the microwave field. By tuning the microwave frequency to the transitions $i = 1, 4,$ or $7$ we are sensitive to the $\sigma-$, $\pi$, or $\sigma+$ component of the microwave magnetic field, respectively, via the corresponding transition's Rabi frequency $\Omega_i$.

Imaging of the DC magnetic C-field was conducted using Ramsey measurements, where the microwave interaction is achieved by *two* $\pi/2$ microwave pulses separated by a Ramsey time $dt_R$, during which neither light nor microwave are applied. The detected variation in optical density as function of $dt_R$ is described by

$$\Delta OD = A - B \exp\left(-\frac{dt_R}{T_1}\right) + C \exp\left(-\frac{dt_R}{T_2}\right) \sin(\delta\, dt_R + \phi) \qquad (2)$$

Here the fit parameters are $A, B, C, T_1, T_2, \delta,$ and $\phi$, analogue to Eq. (1), with $T_1$ and $T_2$ the population and coherence lifetimes, respectively. The Ramsey oscillation frequency $\delta$ is equal to the microwave detuning from the atomic transition. Thus, for an externally fixed microwave frequency injected into the cavity, the value of $\delta$ encodes the C-field amplitude at the position of the sampled atoms, via the Zeeman shift of the selected transition. Using the Breit-Rabi formula [26], the C-field amplitude can then be calculated from $\delta$, and common-mode frequency shift such as buffer-gas shifts can be eliminated when combining measurements on two different transitions.

For construction of spatially resolved images, imaging optics and a CCD camera are used for measuring the transmitted laser intensities. Using a pattern mask at the level of the cavity as well as ray transfer matrix calculation, the imaging optics is found to result in a 1:4 demagnification of the atoms' image on the CCD sensor. In order to reduce noise and limit computation effort on the fitting, the CCD images were binned into image pixels of 3x3 CCD-pixels. Each resulting image pixel corresponds to a 76 μm × 76 μm cross section at the cell and contains the atomic signal integrated within this area. The imaged area has a diameter of 11 mm at the cell, and each image consists of ≈15'000 image pixels. The time series $\Delta OD(dt_{mw})$ or $\Delta OD(dt_R)$ are fitted with equations (1) or (2), respectively, independently for each image pixel. For both Eq. (1) and (2), all seven fit parameters are fitted independently. Examples of such fits to Rabi and Ramsey data of a typical single image pixel are shown in Fig. 3, for data recorded on the $i = 7$ transition. Both data sets are well described by the fit functions of Eq. (1) and (2), respectively. Finally, the fit parameters from each pixel are recombined to obtain images of the physical entities of interest.

Because the measurements are taking place in the time domain, the fit parameters $\Omega$ and $\delta$ of main interest here are largely insensitive to overall variations in the signal amplitude.

## III. EXPERIMENTAL RESULTS

In the following we discuss the images of the DC and microwave magnetic fields across the cell, obtained with the Rabi and Ramsey interrogation schemes. Images obtained generally refer to the x-y plane, but certain information on the field distribution along the z-axis can also be obtained.

### A. Imaging of the DC Magnetic Field

Imaging of the DC magnetic C-field was achieved using the Ramsey interrogation method (see section II-C). Figure 4 shows the C-field amplitude $B_{dc}$ and $T_2$ lifetime images, obtained from the corresponding fit parameters of Eq. (2). Data was recorded with the microwave frequency tuned close to the i = 2 transition and common-mode frequency shifts were removed by using similar measurements on the i = 6 transition. The C-field amplitude is found to be very homogeneous, with a peak-to-peak variation of only 0.13 µT (or 0.3%) over the sampled cell region. Relative uncertainties returned from the fit on each individual pixel are < 0.5% for the C-field amplitude, and < 8% for the $T_2$ time. The images of both the C-field and $T_2$ show a circular symmetry around the same center (indicated by black crosses in the left-hand panels of Fig.4), and do not correlate with any symmetry of the microwave magnetic field distributions measured (see section III-B below). Images of $B_{dc}$ and $T_2$ qualitatively and quantitatively very similar to Fig. 4 are obtained when the microwave frequency is tuned close to the transitions i = 1, 6, and 7. Fitting uncertainties are < 1% for the C-field amplitude and on the level of 2% to 10% for $T_2$ in all cases.

For a more quantitative analysis, we plot the C-field amplitude (Fig. 5a) and $T_2$ (Fig. 5b) as functions of distance r from the center of symmetry (black crosses in the left-hand panels of Fig. 4), for transition i = 2. The C-field dependence on r is well-described by

$$B_{dc}(r) = C_0 + C_2 \cdot r^2 + C_4 \cdot r^4 \qquad (3)$$

shown as fit in Fig. 5a. As expected, the C-field profiles measured on transitions $i = 1, 6$, and 7 show a very similar behavior: All 4 transitions give a consistent value of $B_{dc}(r=0) = C_0 = 40.31$ µT, only slightly below the 40.41 µT calculated from the solenoid geometry and applied current. Because the clock transition ($i = 4$) shifts in second order only with the magnetic field, determination of the C-field amplitude from this transition gives much bigger uncertainties, with data scatter on the level of 0.1 µT, which completely masks the C-field structure visible in Fig. 5a.

The measured C-field amplitudes and $T_2$ times are clearly correlated, as seen from Fig. 4. This can be attributed to inhomogeneous dephasing due to spatial gradients in the C-field, a process well-known from nuclear magnetic resonance (NMR) spectroscopy: at large $r$ values, the Rb atoms diffusing through the buffer gas sample spatial regions of more pronounced differences in C-field amplitude, thus resulting in reduced $T_2$ times [27], [28]. Figure 5b indeed shows a general linear decrease of $T_2$ with $r$ (dashed black line in Fig. 5b), with a slope $\partial T_2/\partial r = -41.6$ µs/mm, which is consistent with a dependence of $T_2$ on the field gradient $\partial B_{dc}/\partial r$ for a C-field distribution governed by the quadratic term $C_2 \cdot r^2$, as is the case here. One further observes that $T_2$ values above (below) the linear trend are measured in regions where $\partial B_{dc}/\partial r$ is locally smaller (larger) than the derivative of the general polynomial of Eq. (3), marked by dashed (solid) vertical arrows in Fig. 5a and 5b. The radial $T_2$ profiles measured on transitions $i = 1, 6$, and 7 again show the same general behavior as found for $i = 2$, however residual deviations from the linear trend are slightly stronger on $i = 1$ and 7 (recorded using the $\sigma^-$ and $\sigma^+$ microwave magnetic field components) than for the $i = 2$ and 6 case (recorded using the $\pi$ microwave field component). This indicates a dependence of the inhomogeneous dephasing on the $m_F$ quantum numbers or magnetic field component involved.

Under conditions of inhomogeneous dephasing in spin echo NMR and for linear and quadratic spatial variation of the static magnetic field the overall transverse relaxation time $T_2^*$ is given by [28], [29]

$$T_2^{*-1} = T_{2,0}^{-1} + \tfrac{1}{3} D \left(G \gamma t_R\right)^2 = T_{2,0}^{-1} + \eta G^2 \qquad (4)$$

Here $T_{2,0}$ is the $T_2$ time at vanishing field gradient, $D$ the diffusion constant of Rb atoms in our cell, $\gamma$ the atoms' gyromagnetic factor, $t_R$ the average Ramsey time, and $G$ the local gradient of the static magnetic field of relevance here. For more complicated field distributions, numerical approaches are generally used [30]. The solid black line in Fig. 5b gives a fit of $T_2^*(r)$ according to Eq. (4) to the measured distribution $T_2(r)$, using $G(r)=dB(r)/dr$ calculated according to the fit of Eq. (3) shown in Fig. 5a, which yields $T_2^*(r=0) = T_{2,0} = 445.5(6)$ µs and $\eta = 1.794(8)$ mm²·µs⁻¹·µT⁻² for the only two free parameters of this fit. Although derived for the case of spin echo, this curve reproduces very well the general shape of the measured $T_2(r)$ dependence. The blue dotted curve in Fig. 5b gives the profile $T_2^*(r)$ calculated for our cell conditions, using the same $T_2^*(r=0) = 445.5$ µs and $\eta = 1.2$ mm²·µs⁻¹·µT⁻² (calculated from $D=7.1$ cm²/s for our cell conditions and $t_R = 0.8$ ms, no free parameters). We attribute this residual difference in $\eta$ to the fact that i) Eq. (4) is derived for spin echo experiments while our measurements are obtained using the Ramsey scheme with varying $t_R$, and ii) the actual field gradient $G(r)$ might be underestimated by neglecting contributions from $dB(r)/dz$ here. The results still indicate that the observed variation $T_2(r)$ can indeed be caused by inhomogeneous dephasing due to magnetic field gradients.

Figure 5c shows the $T_2$ time measured on the $i = 4$ transition. Again, $T_2$ shows a linear dependence on $r$, with a slope $\partial T_2/\partial r = -13.3$ µs/mm, approximately 3 times lower than observed on $i = 1, 2, 6$, and 7. The mean value of $T_2 = 1.5(2)$ ms is considerably higher than the one observed on the other transitions and is in agreement with the intrinsic clock transition linewidth measured for this cell [31].

B. *Imaging of the Microwave Field*

Imaging of the microwave magnetic field inside the cavity was performed using the Rabi interrogation scheme (see section II-C). The microwave transitions i = 1, 4, and 7 are exclusively driven by the $\sigma^-$, $\pi$, and $\sigma^+$ components of the microwave magnetic field, respectively. By tuning the microwave frequency to each of these resonances in turn and extracting the corresponding Rabi frequencies $\Omega_i$ from fits of Eq. (1) to each pixel's data we can therefore determine the amplitude of the magnetic microwave field components as:

$$B_- = \frac{1}{\sqrt{3}} \frac{\hbar}{\mu_B} \Omega_1$$
$$B_\pi = \frac{\hbar}{\mu_B} \Omega_4 \quad (5)$$
$$B_+ = \frac{1}{\sqrt{3}} \frac{\hbar}{\mu_B} \Omega_7$$

where $B_-$, $B_\pi$, and $B_+$ are the amplitudes of the $\sigma^-$, $\pi$, and $\sigma^+$ microwave field components, respectively [21]. Here, the microwave magnetic field amplitudes are defined with respect to their Cartesian basis as [32]:

$$\begin{aligned} B_\pi &= B_z \\ B_- &= \tfrac{1}{2}(B_x + iB_y) \\ B_+ &= \tfrac{1}{2}(B_x - iB_y) \end{aligned} \quad (6)$$

The results obtained for the different microwave field components are shown in Fig. 6, together with the corresponding Rabi oscillation lifetime $\tau_2$. The variations of the microwave magnetic field $\pi$-component are less than 20% of its peak value across the entire image, much stronger than those detected for the C-field and not correlated with the latter. The maximum values for the $\sigma^+$ and $\sigma^-$ components are on the level of 10% and 20%, respectively, of the $\pi$-component's maximum value, which confirms a microwave magnetic field predominantly oriented parallel to the z direction [13]. Note that during clock operation, these $B_-$ and $B_+$ field components can in principle induce AC Zeeman shifts of the atomic levels and thus frequency shifts of the clock (in analogy to AC Stark shift introduced by the optical radiation). However, in clock operation the microwave frequency will be resonant with the i = 4 $\pi$-transition, and thus far off-resonant with respect to the s transitions (by approximately $\geq$ 30 linewidths here), thus AC Zeeman shifts from the $B_\pi$ component can be expected to be largely dominant. The off-center position of the field maximum for the $\pi$-component can be attributed to a shift of the microwave field distribution caused by the cell reservoir located on the upper left rim of the cavity (see Fig. 1b) [33]. Fitting uncertainties for the microwave field amplitude are < 8 nT or < 3% for all three field components.

Similar to the observations made in [23] and on the C-field images (section III-A) we also here see a reduction of Rabi oscillation lifetime, i.e. inhomogeneous dephasing, but this time as a function of the microwave magnetic field amplitude: for each field component, $\tau_2$ reaches its maximum in regions where the gradient of the microwave magnetic field amplitude is small, while $\tau_2$ is lower in regions of high microwave magnetic field gradients. Fitting uncertainties for $\tau_2$ are generally below 5%, reaching up to 8% only in the region of highest $\tau_2$ in the image of the $\pi$ component (due to the relatively short maximum $dt_{mw}$ employed here).

In the microwave magnetic field data of Fig. 6a a small region close to the lower left edge of the imaging area exhibits some deviations from the general structure seen. We interpret this structure as an artefact, arising from the unfavorable combination of low magnetic field values (and thus long Rabi oscillation period) and short Rabi oscillation lifetimes $\tau_2$. Such limitations can be overcome by performing the measurements at higher microwave power levels that enable again observation of several Rabi oscillations within the $\tau_2$ time.

In order to assess the detection limit of the method, measurements were repeated on transition $i = 4$ for different levels of microwave power sent into the cavity, using a photodiode as detector here. The inset of Fig. 7 shows that the square of the measured microwave field amplitude scales linearly with the microwave power sent to the cavity, as expected. Clear Rabi oscillations can still be observed for microwave field amplitudes as low as 0.05 µT. The signal-to-noise ratio (Rabi oscillation amplitude $C$ over rms data scatter) found for the Rabi data of Fig. 3a, although taken on a single image pixel only, is even better than that of Fig.7, and the difference in magnetic field amplitude detected is mainly reflected by the different time scale of the Rabi oscillations – note the difference in x-axis scaling by a factor of 10. We therefore conclude that also in the imaging mode microwave magnetic field amplitudes of 0.05 µT or below can be detected with this method.

In principle, the images of the microwave magnetic field shown in Fig. 6 give for each image pixel an average value of the field amplitude, integrated over the 25 mm length of our thick cell along the laser beam propagation direction (parallel to the z-axis), which is a conceptual difference to the work in [23] on a thin cell. One can however still obtain some statistical information on the field distribution along the z-axis here, by calculating the Fourier transform spectrum of an image pixel's ΔOD data (see Fig. 3), inspired by similar techniques applied in NMR and Fourier transform spectroscopy. We illustrate this approach for the example of the microwave field π component (Fig. 6b), but it is equally applicable to the other field components or the C-field data of Fig.4. Figure 8 shows the Fourier transform magnitude for pixels (88,98) and (39,15) of Fig. 6b, that are from regions of high and low $B_\pi$ amplitude, respectively, as well as for the data obtained by integrating ΔOD over the entire image (x and y coordinates) for each value of $dt_{mw}$. The Fourier frequency axis is then scaled into microwave magnetic field amplitude according to Eq. (5).

The resolution of the Fourier spectrum in Fig. 8 is ≈ 0.4 kHz, or 30 nT, due to the limited length of the $dt_{mw}$ time-domain signal used here. Nevertheless all three Fourier spectra show narrow peaks for the $B_\pi$ distribution, with a spread of 10% or less for the two individual pixels. At low frequencies the spectrum diverges due to the exponential background in the time-domain signal, which is of no interest here and therefore is omitted from Fig. 8. For pixel (39,15), the small peak at $B_\pi < 1$ µT indicates the presence of some z-region with very low field amplitude. The integrated signal shows a structure composed of the distinct features of both extreme pixels selected, plus an intermediate feature at 4.5 µT, as expected from the general field distribution in Fig. 6b. In the integrated signal small dips are seen at 4.14, 4.3, and 4.55 µT, making this signal appear to be constituted of several distinct peaks. Given the overall smooth field distribution of Fig. 6b, these dips might well be artifacts without statistical significance, arising from measurement noise or instabilities converted by the FFT routine employed. Note that the contribution of the low field amplitudes like in pixel (39,15) to the integrated signal is small, due to the small image area in Fig. 6b showing such low amplitude values. We thus conclude that also the variation of $B_\pi$ along the z-axis is small (< 10%), otherwise a more pronounced broadening of the $B_\pi$ distribution should be observable for all 3 traces in Fig.8.

*C. Optical Pumping Efficiency*

Figure 9 shows the image of the *A* parameter in Eq. (1), which is a measure of the optical pumping efficiency, for the microwave π component of Fig. 6b here. The fringes seen are rather strong, with a small-scale variation by a factor of 4, probably due to optical interferences caused by not perfectly planar cell windows – notably at the upper left edge where the cell reservoir is located – and by the apertures of the imaging optics. These fringes are not observed in the microwave field images of Fig. 6, and only slightly in the $\tau_2$ images and fitting uncertainties, which demonstrates the robustness of the method against variations in light intensity. Also in the Ramsey scheme the impact of these fringes is very small, with only small distortions visible in the $T_2$ images (see Fig. 4).

## IV. Conclusion

We have employed imaging techniques based on time-domain Ramsey and Rabi spectroscopy of Rb atoms for measuring experimentally the DC and microwave magnetic field distributions inside a compact microwave resonator holding a Rb vapor cell. The π-component of the microwave magnetic field – relevant for clock operation – is found to vary by less than 20% of its maximum value over the entire imaging region. This level of homogeneity is expected to be sufficient for the observation of high-contrast Ramsey fringes in pulsed clock operation [34]. The $\sigma^+$ and $\sigma^-$ field components – orthogonal to the π-component – show much lower peak amplitudes of 10% and 20% of the π-component's maximum value, respectively. Using Fourier Transform analysis of the signals, information on the otherwise unresolved field distribution along the z-axis (i.e. propagation direction of the laser beam) can also be obtained and indicates a very homogeneous distribution of the microwave π-component along the z-axis, also favorable for pulsed clock operation. Imaging of the C-field amplitude in the cell reveals <1% amplitude variation across the cell, and along with $T_2$ relaxation times that correlate with the gradient of the C-field amplitude, due to inhomogeneous dephasing. Fourier analysis as demonstrated for the microwave fields can also be applied to the C-field data.

The employed imaging techniques allow assessing experimentally the DC ("C-field") and microwave magnetic field distributions in a vapor-cell atomic clock physics package, under real operating conditions of the fully assembled cell and resonator package. This possibility is of particular importance for the microwave magnetic field whose distribution generally is affected by the presence of the dielectric cell material. Given the relatively limited fabrication tolerances possible for such cells, such in situ assessment of the microwave field distribution is of high relevance for the development of compact high-performance vapor-cell atomic clocks or other atomic sensors. The method can also be conveniently used as a diagnostics tool for detecting changes in C-field amplitude after extended clock operation time, due to re-magnetization or hysteresis of the magnetic shields employed.

ACKNOWLEDGMENT

We thank A. K. Skrivervik and A. Ivanov (Laboratory of Electromagnetics and Acoustics (LEMA), École Polytechnique Fédérale de Lausanne (EPFL), Lausanne, Switzerland) for many helpful discussions and support on the microwave cavity, M. Pellaton (UniNe-LTF) for making the vapor cell, and P. Scherler (UniNe-LTF) for support on design, manufacturing and assembly of the microwave cavity and its physics package.

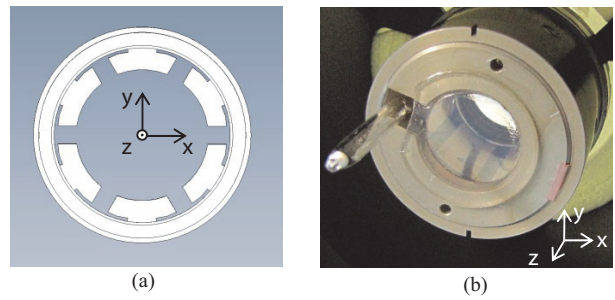

Figure 1: a) Cross-section drawing of the microwave cavity with its electrode structure. The coordinate system used throughout the paper is shown; the z-axis coincides with the cavity's axis of cylindrical symmetry. b) Assembled microwave cavity with the Rb vapor cell. For better visibility, the coordinate system is offset from its true origin here.

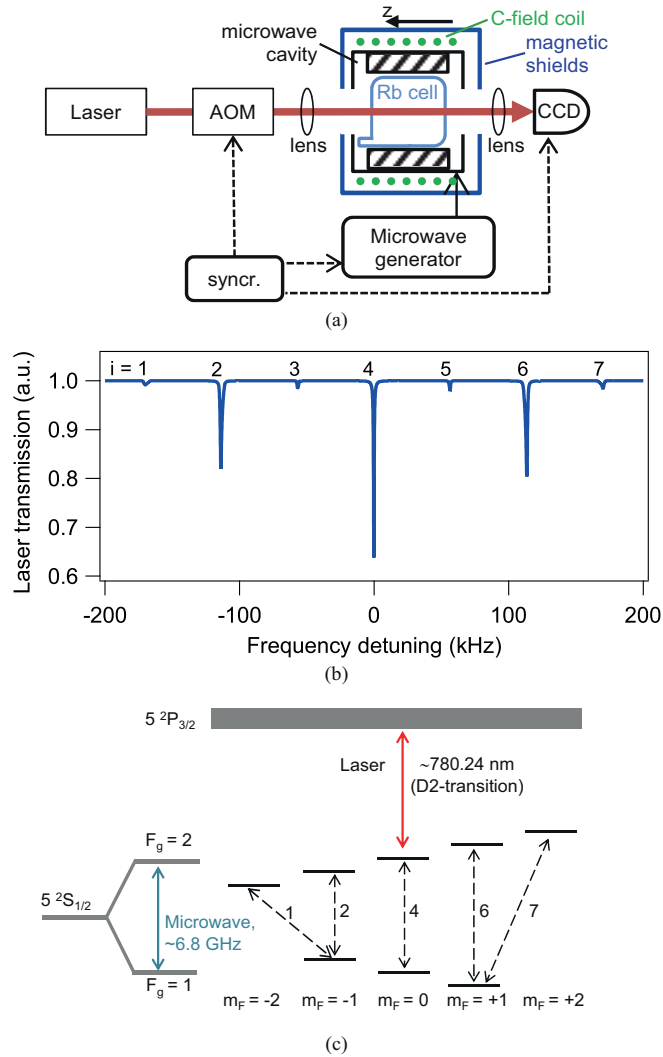

Figure 2: a) Experimental setup used for field imaging. b) Double-resonance optical transmission signal recorded in cw operation, showing the Zeeman-splitting of the microwave ground-state transition into 7 lines for a C-field of $B_{dc} = 8.6$ µT. c) Atomic level scheme of the involved $^{87}$Rb atomic states. Dashed black arrows indicate the microwave transitions $i = 1$ to 7. Due to the buffer-gas, the $5P_{3/2}$ excited-state hyperfine structure is not resolved.

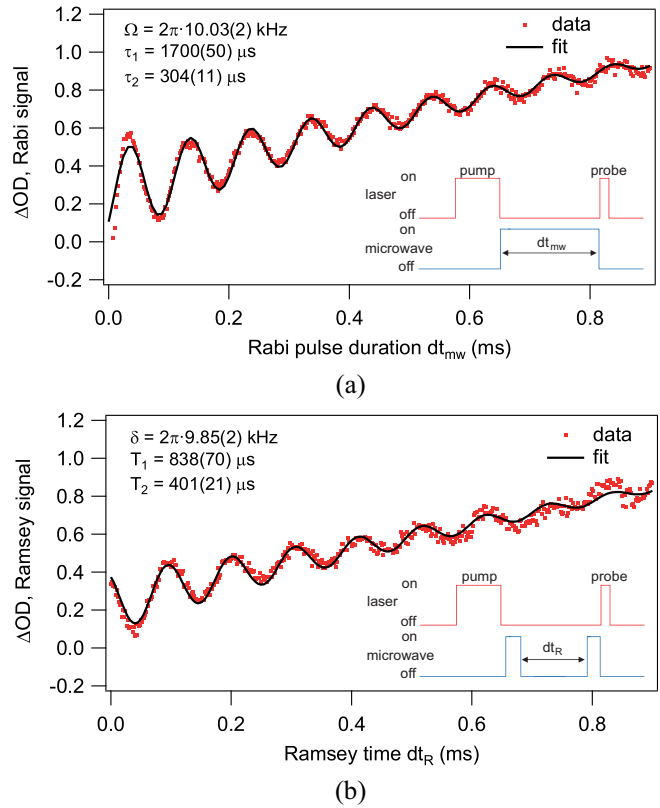

Figure 3: Typical examples of time-domain oscillations in the CCD signal, recorded on the *i* = 7 transition here. a) Rabi data for image pixel (80,80) from Fig. 6c, corresponding to a microwave magnetic field amplitude B = 0.5 µT. b) Ramsey data for an image pixel at the center of symmetry, from data similar to that of Fig. 4. Insets show the pulse sequences employed. In both Figs. 3a and 3b the signals are normalized such that ΔOD= 0 corresponds to the cell's optical density under conditions of optical pumping but without any microwave interaction applied.

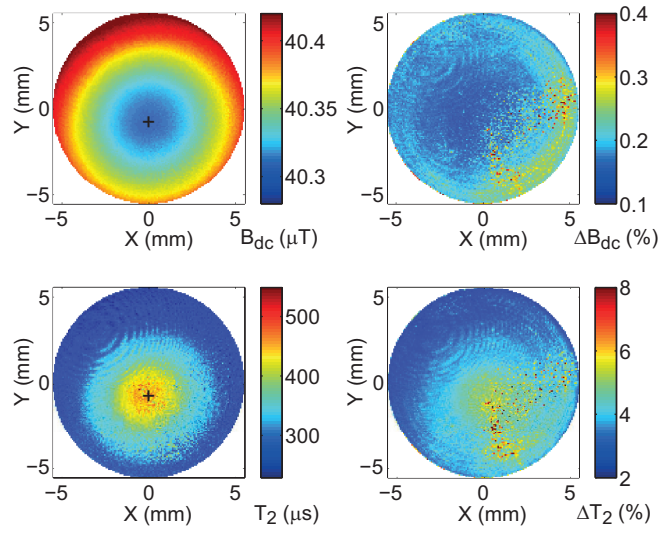

Figure 4: Imaging results measured on transition $i = 2$ ($m_{F=1} = -1 \rightarrow m_{F=2} = -1$). Upper left panel: DC magnetic field amplitude $B_{dc}$. Lower left panel: $T_2$ times. Right-hand panels: corresponding uncertainties returned from the fitting routine. Black crosses in the left-hand panels give the centers of circular symmetry found for both images. See Fig. 1 for definition of the coordinate system used.

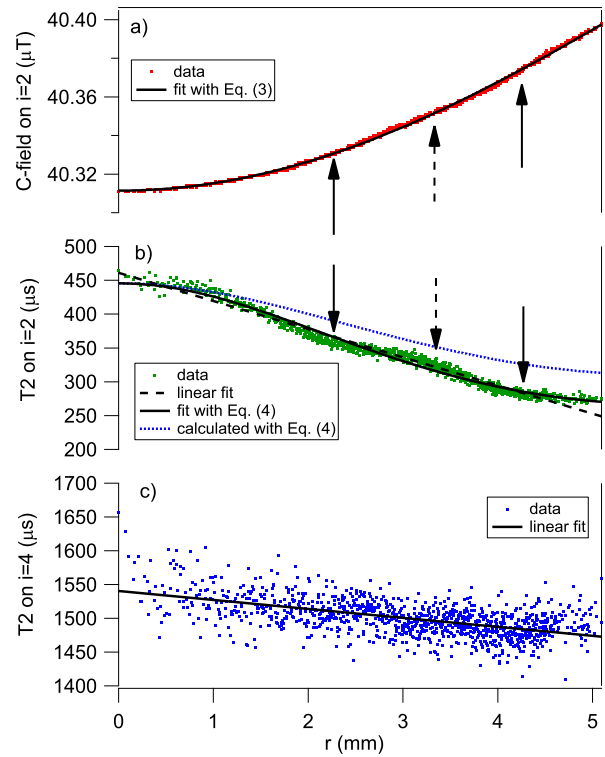

Figure 5: a) radial profile of the C- field amplitude measured on transition $i = 2$ (dots) and $4^{th}$-order polynomial fit to the data (solid line), b) radial profile of the corresponding $T_2$ time, measured on the $i = 2$ transition. Solid arrows mark regions of locally higher c-field gradient and relatively lower $T_2$ times, and dashed arrows mark regions of opposite characteristics. c) radial profile of the $T_2$ time measured on the $i = 4$ transition.

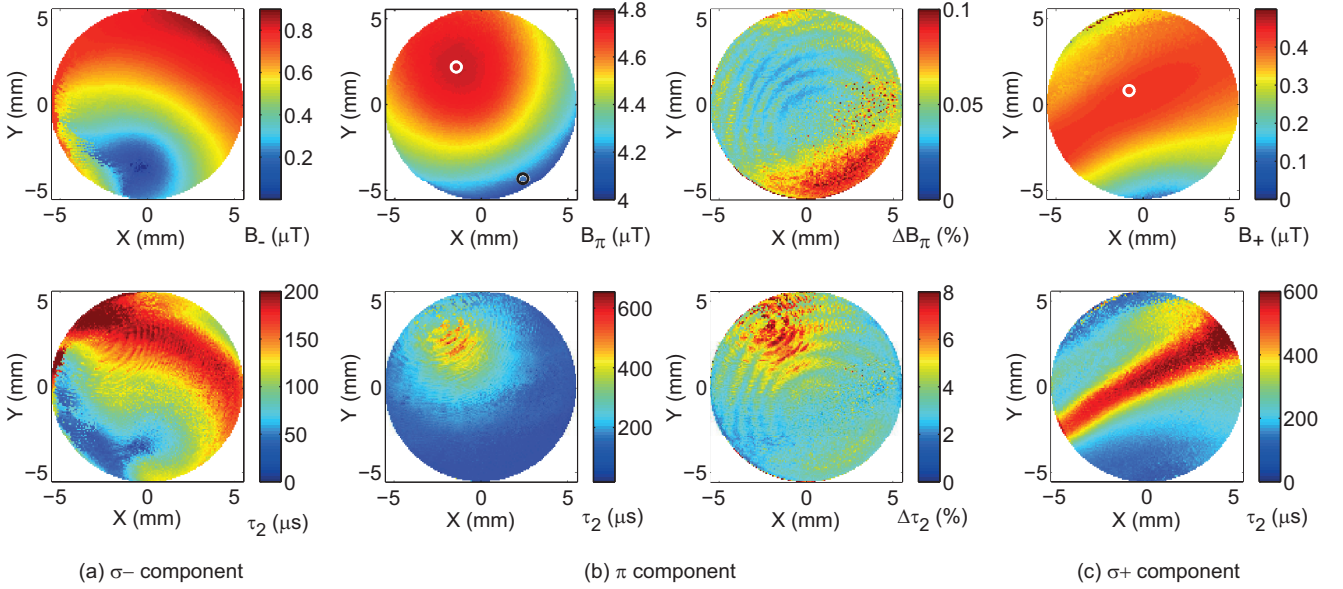

Figure 6: Imaging results for the amplitude $B_-$, $B_\pi$, $B_+$ of the three microwave magnetic field components and corresponding Rabi oscillation lifetimes $\tau_2$. a) $\sigma-$ component, measured on transition $i = 1$ ($m_{F=1} = -1 \to m_{F=2} = -2$); b) $\pi$ component, measured on transition $i = 4$ ($m_{F=1} = 0 \to m_{F=2} = 0$); c) $\sigma+$ component, measured on transition $i = 7$ ($m_{F=1} = +1 \to m_{F=2} = +2$). Note the difference in color scale between the sub-panels. Uncertainties for the microwave magnetic field amplitude and Rabi oscillation lifetimes are shown for the $\pi$ component only (sub-panel b), but are similar for the $\sigma$ components. The white and black circle in sub-panel b indicate the pixels (88,98) and (39,15), respectively, used for the Fourier analysis of Fig. 8. The white circle in sub-panel c indicates pixel (80,80) used in Fig. 3a.

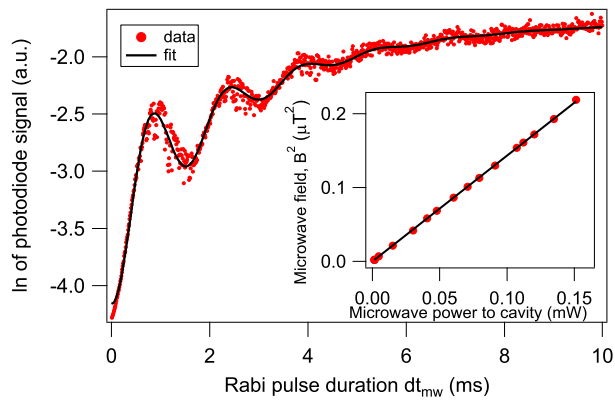

Figure 7: Rabi oscillations can still be clearly observed for a microwave field amplitude as low as $B = 0.047$ µT (microwave power of $P_{mw}=1.5$ µW sent to the cavity). The square $B^2$ of the microwave field amplitude scales linearly with the microwave power $P_{mw}$ sent to the cavity, see inset.

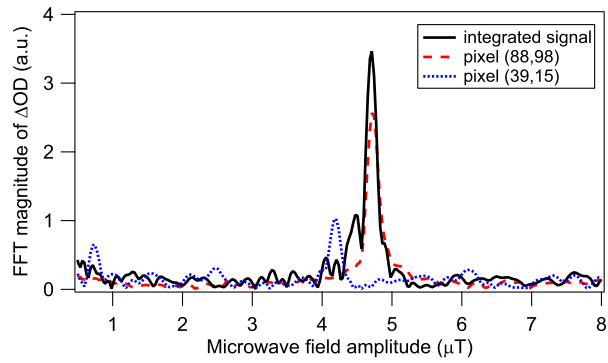

figure 8: Relative contribution of the different field amplitudes for the microwave field π component (see Fig. 6b), obtained by Fourier transform analysis of the ΔOD($dt_{mw}$) signal. Solid black line: signal integrated over the entire image; dashed red line: pixel (88,98) with high $B_\pi$; dotted blue line: pixel (39,15) with low $B_\pi$. The position of the selected pixels is shown in Fig. 6b. For better visibility, the FFT magnitudes of pixel (39,15) and for the integrated signal were multiplied by factors of 2 and 15, respectively, to compensate differences in FFT magnitudes caused by the different $T_2$ times of the time-domain signals.

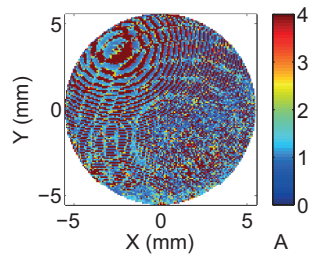

Figure 9: Image of fit parameter A of equation (1) for Rabi images of the microwave magnetic field π component shown in Fig. 6b.